\documentclass[conference]{IEEEtran}
\IEEEoverridecommandlockouts
\pdfoutput=1
\usepackage{cite}
\usepackage{amsmath,amssymb,amsfonts}
\usepackage{algorithmic}
\usepackage{graphicx}
\usepackage{textcomp}
\usepackage{xcolor}
\def\BibTeX{{\rm B\kern-.05em{\sc i\kern-.025em b}\kern-.08em
    T\kern-.1667em\lower.7ex\hbox{E}\kern-.125emX}}
\begin{document}

\title{FOX: Hardware-Assisted File Auditing for Direct Access NVM-Hosted
Filesystems\\
\thanks{\textsuperscript{*}Note: This work was revised three times in 2019 while the author was a PhD student at the UCF.}
}

\author{\IEEEauthorblockN{1\textsuperscript{st} Mao Ye}
\IEEEauthorblockA{\textit{Department of ECE} \\
\textit{University of Central Florida}\\
mye@knights.ucf.edu}
}

\maketitle

\begin{abstract}

With emerging non-volatile memories (NVMs) entering the mainstream market, several operating systems start to incorporate new changes and optimizations. One major OS support is the direct-access for files (DAX) feature, which enables efficient access for files hosted in byte-addressable NVM systems. With DAX-enabled filesystems, files can be accessed directly similar to the memory with typical load/store operations. Despite its efficiency, the frequently used system call, dax-mmap, is troublesome for system auditing. File system auditing is mandatory and widely used because auditing logs can help trace potential anomalies, suspicious file accesses, or used as an evidence in digital forensics. However, the frequent and long-time usage of dax-mmap blinds the operating system or file system from tracking process operations to shared files after the initial page faults. This might results in imprecise casualty analysis and leads to false conclusion for attack detection.

To remedy the tension between enabling fine-grained and whole file system auditing and leveraging the performance of NVM-hosted file systems, we propose a novel hardware-assisted auditing scheme (\textit{FOX}). FOX enables file system auditing through lightweight hardware-software changes which can monitor every read/write access event for mmapped files on NVM. Additionally, we propose the optimized schemes, that enable auditing flexibility for selected files/memory range. By prototyping FOX on a full-system simulator, Gem5, we observe an average performance overhead of only 9 to 6.4\% reduced throughput and 48\% extra writes compared to our baseline. Compared to other instrumentation-based software schemes, our scheme is low-overhead and secure. 

\end{abstract}


\section{Introduction}
To embrace the arrival of Non-Volatile Memory (NVM) products in market, such as Intel's Optane series, major operating system (OS) vendors have readily provided means for integration of such devices\cite{IntelOpt,DAX,WindowsDAX}. The key properties of emerging NVMs that can benefit applications are their low latency and byte-addressability\cite{NVWAL,byteManage,NVMperformance,NVMcache,xu2019finding, caulfield2012providing}. Additionally, NVM can serve as memory, storage or both owing to their dual memory and storage characteristics. 
Files residing on NVM are permanently stored there unless deleted. Accessing NVM-hosted files no longer requires expensive copy operations and buffering of data in the software-managed page cache in memory, which is a new feature called direct access(DAX)\cite{DAX}. Further files can be accessed directly by being mapped into the address space of applications using the {\tt mmap} call. 
DAX allows accessing files efficiently and directly using load/store operations, without expensive copy/buffering operations, which improves performance significantly\cite{xu2017nova}. Furthermore, to achieve high performance, applications only need to memory-map a file and access it following applications' logic, similar to accessing normal memory allocations\cite{XuNova,XuThesis,xu2019finding}.

Unfortunately, the performance advantages achieved by DAX-enabled filesystems come with other consequences. Current systems rely on auditing/monitoring tools to collect large amount of logs daily for multiple purposes\cite{ma2015accurate,lee2013loggc,xu2009detecting}. One reason is that these data
facilitates predicting performance bottleneck\cite{nagaraj2012structured,xu2009detecting}. Another reason is that recording sequences of monitoring/auditing events plays a critical role in reconstructing the dependency graph of processes and files to identify the origin of attacks\cite{ma2017mpi,gao2018aiql,liu2018towards,goel2008reconstructing,hossain2017sleuth,kwon2018mci,king2003backtracking,sitaraman2005forensic,goel2005forensix,king2005enriching,renroot,unearthing}, and provide digital trace evidence on the victim system\cite{malware,krishnan2010trail}. Most intrusion detection/provenance systems consider process, files, sockets, etc. as objects and their interaction as events. Current research either focuses on events collecting, categorizing and filtering\cite{lee2013loggc,ma2015accurate,sitaraman2005forensic}, or on optimization of casualty analysis, attack/malware detection and enabling system recovery to what before the attack entry point\cite{goel2005taser,kwon2018mci,hossain2017sleuth,bates2015trustworthy}, or both. 
Regardless, dax-mmap() is problematic for event recording/analysis because after the system call is recorded, any read or write to the mmapped files is invisible to operating systems. That is, any legal or tainted operation within the mmapped files is not known. Up till now, OS-level monitoring tools are not able to monitor the mmap()-ed region or can only trace the page faults occurred within\cite{tracer}. 
On the other hand, the ext4 that supports dax, does not provide journaling for daxed device, making consistency of a dax-style mmaped file becomes a programmer's duty rather than that of the filesystem itself\cite{xu2019finding,xu2017nova}. Hence, file system is unaware of operations in mmap() region either. Another side effect is that loss of track renders snapshot creation difficult for DAX-supported files systems, impairing the system reliability. Thus, unless efficient monitoring solution is available, NVM-hosted filesystem has the limitation to be deployed on critical systems or for large-scaled distributed service where security regulation, forensic trail, and reliability are of paramount importance. Such systems include big data web service and database, where DAX support is useful and promising to improve performance\cite{keeton2015machine,arulraj2019data}.

In this paper, we want to provide a hardware-assisted file access monitoring scheme that enables fine-grained access records for files that are mmapped on NVM. There are software approaches that can track the store orders in mmapped region by triggering additional page faults\cite{XuNova,kateja2019tvarak,kateja2019lazy}. However each fault induces a latency of micro seconds which is about 3 orders of magnitude of a memory access\cite{XuNova}. Other system call-level monitoring tools cannot observe operations within mmapped regions. We notice that memory trace tools, i.e. PIN, Valgrind, can be used to get detailed read/write order, and can trace selected file accesses, but they usually come with instrumentation and page unmapping, that significantly degrades application performance\cite{nethercote2007valgrind,reddi2004pin}. 
In contrast, hardware-assisted memory tracing is faster, un-distorted, completed but is un-customized and complicated to analyse\cite{baomtt,grimsrud1993bach,TraceARM}. Ideally, we would like to provide a solution that achieves the low overhead of hardware approaches but with software-like flexibility.

To  this  end,  we  propose  a  novel  scheme to enable monitoring  in  NVM-hosted   file-system through hardware-software co-design. In particular, we propose FOX, a novel scheme that relies on minimal changes at the memory controller  and  the  modification  on  Linux  kernel,  to  enable monitoring inside mmap()-ed region in  NVM-hosted  systems. The solution itself is not NVM-specific and can be generally applied.

\indent{}Our hardware/software co-design scheme supports monitoring mmap()-called file accesses and modifications in DAX-enabled NVM-aware filesystem. Fox uses page fault handler to modify the physical address to encode the monitoring metadata. Users have options to choose from different monitoring schemes according to their own needs. 
The major scheme selectively monitors data in NVM which is accessible by dax-mmap function. But to monitor write or read requests, or both, is optional. Further FOX allow users to monitor files only under a specified directory, a common option that has been supported in the software world to address secrecy priority. The persistent memory and directory-selected monitoring shows an acceptable and minor overhead. We evaluate our framework on counter-based encrypted NVM as advocated\cite{SS}. However, our schemes are well suited to non-secure NVM in no effort of change.

To the best of our knowledge, our paper is the first to provide a fine-grained monitor scheme to fill in the blank in security protection for NVM-aware filesystem and for general provenance systems\cite{bates2015trustworthy}.\\
\indent{} Our contributions in this paper are:
\begin{itemize}
    \item Our paper is the first to observe and discuss the challenges of implementing filesystem monitoring in NVM systems.
    \item We propose a hardware/software co-design scheme, \textit{FOX}, to support fine-grained file access monitoring in NVM-hosted filesystems.
    \item We discuss the different design options and optimizations to enable practical file system monitoring for NVM.
 
    \item We prototype our scheme through modifying a recent Linux kernel and run it through a full-system simulator that models our hardware changes.
    
\end{itemize}

The rest of this paper is organized as follows. Section \ref{sec:back} provides relevant background. Section \ref{sec:design} describes the design flow and the implementation of Fox. Section \ref{sec:method} introduces the evaluation methodology. Section \ref{sec:eval} discusses the experiments' results. Section \ref{sec:rela} describes related work. Eventually, Section \ref{sec:con} concludes the paper. 




\section{Background}
\label{sec:back}
Emerging Non-Volatile Memories (NVMs) promise merging storage and memory systems through providing a fast, non-volatile and dense devices\cite{PCM3,PCM3,PCM2}. Recently, Intel and Micron announced the 3D XPoint memory technology which has latency comparable to DRAM while featuring high-density, low idle power and non-volatility. Emerging NVMs are expected to be attached to the memory bus and able to host both memory pages and filesystem data. To enable efficient access to NVM-hosted filesystems, the direct-access (DAX) feature is supported by many OSes.

For the rest of this section, we discuss the DAX support, the current internal monitoring tools and the impact of DAX-mmap on monitoring support. 
\subsection{Direct-Access (DAX) Support}
Filesystems have always been evolving to optimize their performance for emerging storage devices \cite{EuroSys14}. The advent of fast and byte-addressable NVMs shows that, for this type of media, generic block layer and copying data between storage and OS page cache significantly degrade system performance\cite{Moneta,condit2009, SCMFS}. The main reason is that current filesystems require double-copy of data passing between user application and storage: one from storage to page cache, and the other from page cache to user application buffer. Moreover, double-copy sits in critical I/O path of both read and write operations, which unavoidably affects system performance. Thus, it is more efficient to attach NVM directly to memory bus and map their addresses to user applications virtual addresses. 
Hence in NVM-hosted filesystems, direct-access is provided to bypass page cache\cite{xu2017nova,SCMFS,condit2009,EuroSys14,DAX}. The direct-access function in Ext4 is called DAX, which is an optimized version to replace XIP, an early function with similar purpose \cite{DAX}. An overview of the DAX support provided in filesystems is shown in Fig (\ref{fig:DAXFS}).

\begin{figure}[htbp]
    \label{fig:DAXFS}
    \vspace{-1em}
    \centerline{\includegraphics[scale=0.25]{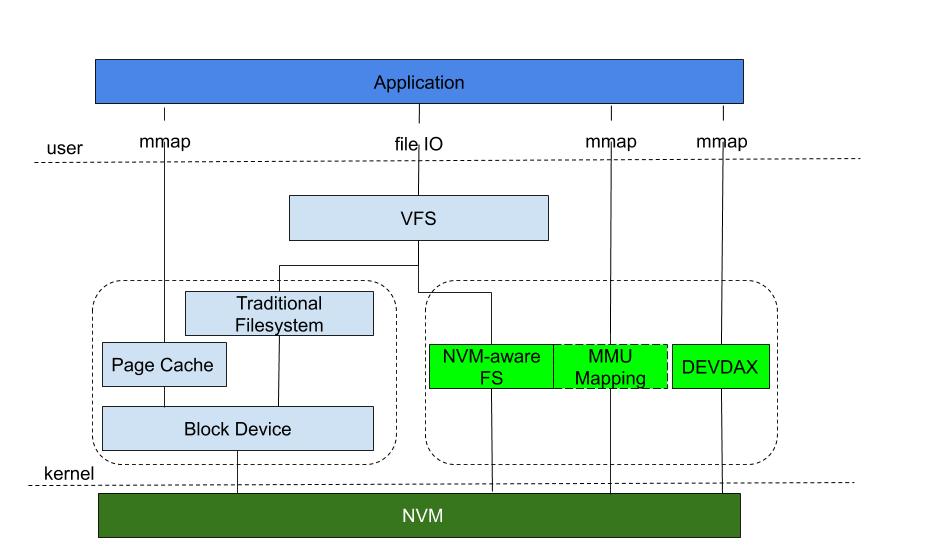}}
    \caption{DAX Support in Filesystem.}
    \vspace{-0.8em}
    \label{fig:DAXFS}
\end{figure}

As shown in the Figure 1, DAX support would directly memory map some physical addresses of a file on NVM into some address range of a user space. Thus, any access to that file will be similar to accessing the memory; virtual address will be converted to physical address then the access completes as read/write to the resulting physical address.

\begin{figure*}[htbp]
\label{fig:depend}
\centerline{\includegraphics[scale=0.3]{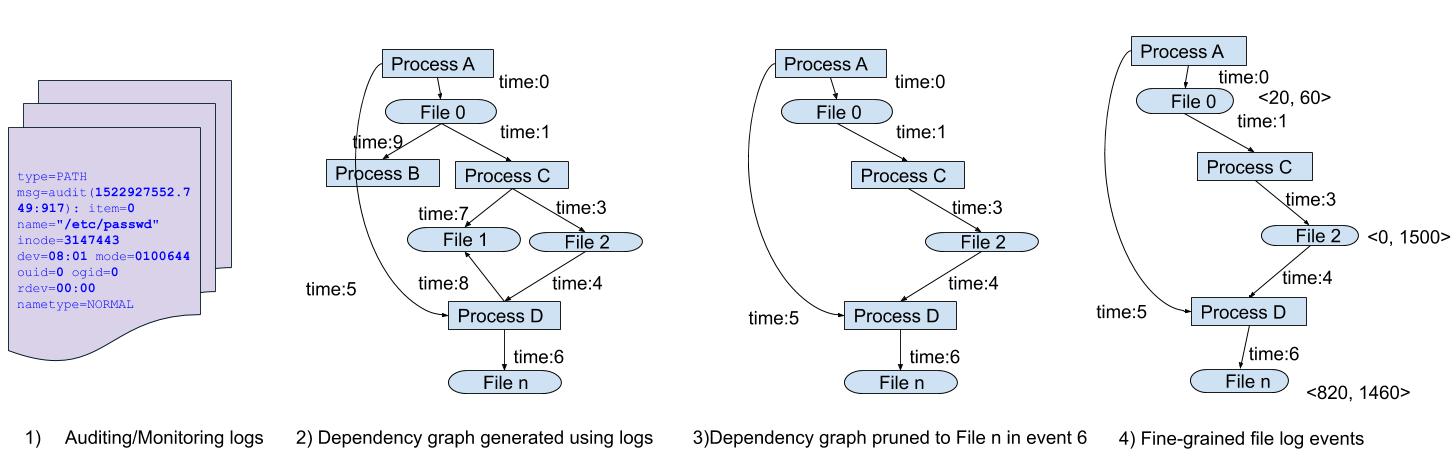}}
\caption{Using audit logs to construct dependence graph }
    \label{fig:windowsMon}
\end{figure*}

\subsection{Auditing/Monitoring logs and casualty analysis}

Auditing is a major requirement for systems that are subject to digital forensics, strict auditing level conditions and/or systems require timely detection of anomalies. For instance, the National Institute of Standards and Technology (NIST) defines a set of Audit and Accountability guidelines for federal information systems and organizations, where it describes cases on which organizations might audit each file access (both successful and unsuccessful)\cite{NIST}. Audit subsystem is nearly mandatory to mainstream operating systems,i.e Linux audit framework{}. However, the file access events still remains to be of coarse-grainularity. 

Additionally, monitoring are widely deployed by internal and external tools.The most widely used system performance monitoring tool might be sysinternal series from Windows\cite{Sysinternals}. Its legacy Filemon(different from FileMon on UNIX) has recently been incorporated into Promon while it maintains offset-and-data-length level read/write file access monitoring for each process. Other mature monitoring tools include ETW for Windows, DTrace for Mac OS\cite{cantrill2004dynamic,ETW}. Inotify was once believed as a completed solution for reliably monitoring filesystem event in Linux\cite{inotifylimitation}. Simply using a file descriptor, inotify can monitor changes of files, directories. Therefore, it is widely adopted by systemtap, osquery, watchman and so on, all are commonly-seen monitoring applications\cite{Systemtap,osquery,watchman}.
However, one of its limitations that mmap() of a file is not notified will cause significant monitoring failure for DAX-mmapped applications.

As a matter of fact, system-call  monitoring  and audit logs have been widely studied. With audit logs, we can trace back the first entry of attacks, and reveal how the attacks are ramified\cite{ma2017mpi,gao2018aiql,liu2018towards,goel2008reconstructing,hossain2017sleuth,kwon2018mci,king2003backtracking,sitaraman2005forensic,goel2005forensix,king2005enriching,renroot,unearthing,bates2015trustworthy}.
Casualty analysis is employed to generate provenance dependence graph based on auditing/monitoring data, as shown in Figure 2. 
The dependence graph is reconstructed automatically using a sequence of events depending on a user-specified event or from multi-hosts events\cite{king2003backtracking,king2005enriching}. The event here must contain at least 3 items, $<$object, operation, object$>$, where objects could be processes, files and operations could be read/write/delete/execute etc. Casualty analysis can detect file system intrusion and it has been shown that fine-grained casualty analysis can help to solve dependency explosion problem. Fine-gained file access offset therefore is useful in providing dependency evidence, such as shown in Figure 2\cite{sitaraman2005forensic}. Recent researches have made efforts to improve dependencies and analysis accuracy, such as taking advantage of machine learning algorithms, devising new models, using domain-specific knowledge and smartly reducing irrelevant data\cite{ma2017mpi,gao2018aiql,liu2018towards,hossain2017sleuth,kwon2018mci,bates2015trustworthy}.

\section{Design}
\label{sec:design}
\subsection{Motivation}
As mentioned before, we notice that the performance enhancer dax-mmap() will hide the fine-grained dependencies of events if multiple processes access an NVM-resident file for a long time, due to the loss of track of file operations after the initial page faults. This poses a security vulnerability for NVM-support file system for detecting attacks and forensic investigation. Therefore, we propose FOX, a hardware-assisted framework to audit file accesses. This section first presents the threats for current NVM-aware filesystem, followed by design schemes for FOX in details and the relevant discussion.

\subsection{Threat Model}
Our threat model consider threats including insider/network attack, multi-stage attack, advanced targeted attacks etc. presented in other system provenance research works \cite{ma2017mpi,gao2018aiql,liu2018towards,goel2008reconstructing,hossain2017sleuth,kwon2018mci,king2003backtracking,sitaraman2005forensic,goel2005forensix,king2005enriching,renroot,unearthing}. Linux Kernel is trusted in our setting. The kernel level threats, such as data-driven attack, kernel malware or other compromises that can modify the monitoring/auditing data, kernel configuration are beyond the scope of this work\cite{xiao2015kernel,rhee2011characterizing,lanzi2009k} 

\subsection{General Picture of Fox}
Fox, as a whole, serves a fine-grained file access auditing framework for NVM-supported filesystem. It manages monitoring entry space, checks specific monitor requirement, issues monitoring requests and ensures strict orders for application memory access and monitoring access. \\
\begin{figure*}[htbp]
 
   \centerline{\includegraphics[
width=0.95\textwidth, height=5cm]{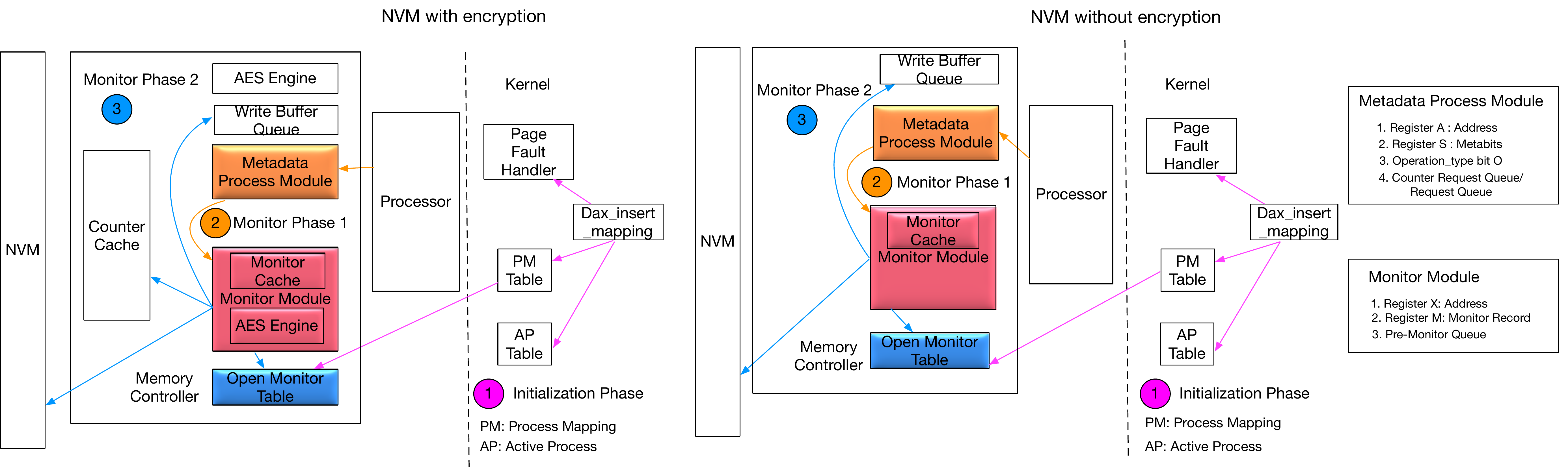}}
    \caption{Monitor process in Fox}
    \label{fig:Fox}
    
\end{figure*}
\indent{}The Fox design requires a lot of coordination for both hardware and software. Memory controller, as the major hardware player, is modified to be in charge of metadata recognition and monitoring issue, while  Linux Kernel comprehends the monitoring metadata from the processes, and communicates the metadata with memory controller accordingly. The whole monitoring process consists of three phases: \textbf{Initialization phase}, \textbf{Monitor phase 1}, and \textbf{Monitor phase 2}. \\
\indent{} In \textbf{Initialization phase}, any process that accesses mmapped files in NVM device first time will be given auditing metadata in their physical addresses when page faults are triggered for accessing monitored files. Meanwhile, the file path, process id and other information will be updated into a hardware table, served as a monitoring entity.  Once the physical address is acquired for each request, memory controller will take over the \textbf{Monitor phase 1} in which any physical address from a request will be analyzed to identify whether it contains auditing metadata. If the answer is yes, the request with that address with be passed to \textbf{Monitor phase 2} to prepare audit record and an audit logging request will be issued. The whole scheme is shown in Figure \ref{fig:Fox} and will be discussed in detail later. Fox successfully achieves the goal of selective monitoring by user-defined rules and only incurs a small percentage of extra writes.

Like software-based monitoring, FOX has specific monitor contents and monitor log management, but first of all let me discuss the most challenging part in the design, that is how to communicate monitoring metadata between software(kernel) and hardware (memory controller). 

\subsection{Software and hardware interaction}
Hardware lacks the ability to understand high-level semantics, therefore, we encounter many problems in supporting hardware monitoring. 

First, how would the memory controller know if the address being read/written should be monitored or not? Second, how does the memory controller know the file ID, User ID and other attributes need to be logged in the monitor entry? Third, how would the memory controller know where to write the monitor logs, e.g., the index of the circular buffer? All these questions are to be answered in this work.
\begin{figure}[htbp]
    \label{fig:Metabit}
    \vspace{-0.8em}
    \centerline{\includegraphics[scale=0.28]{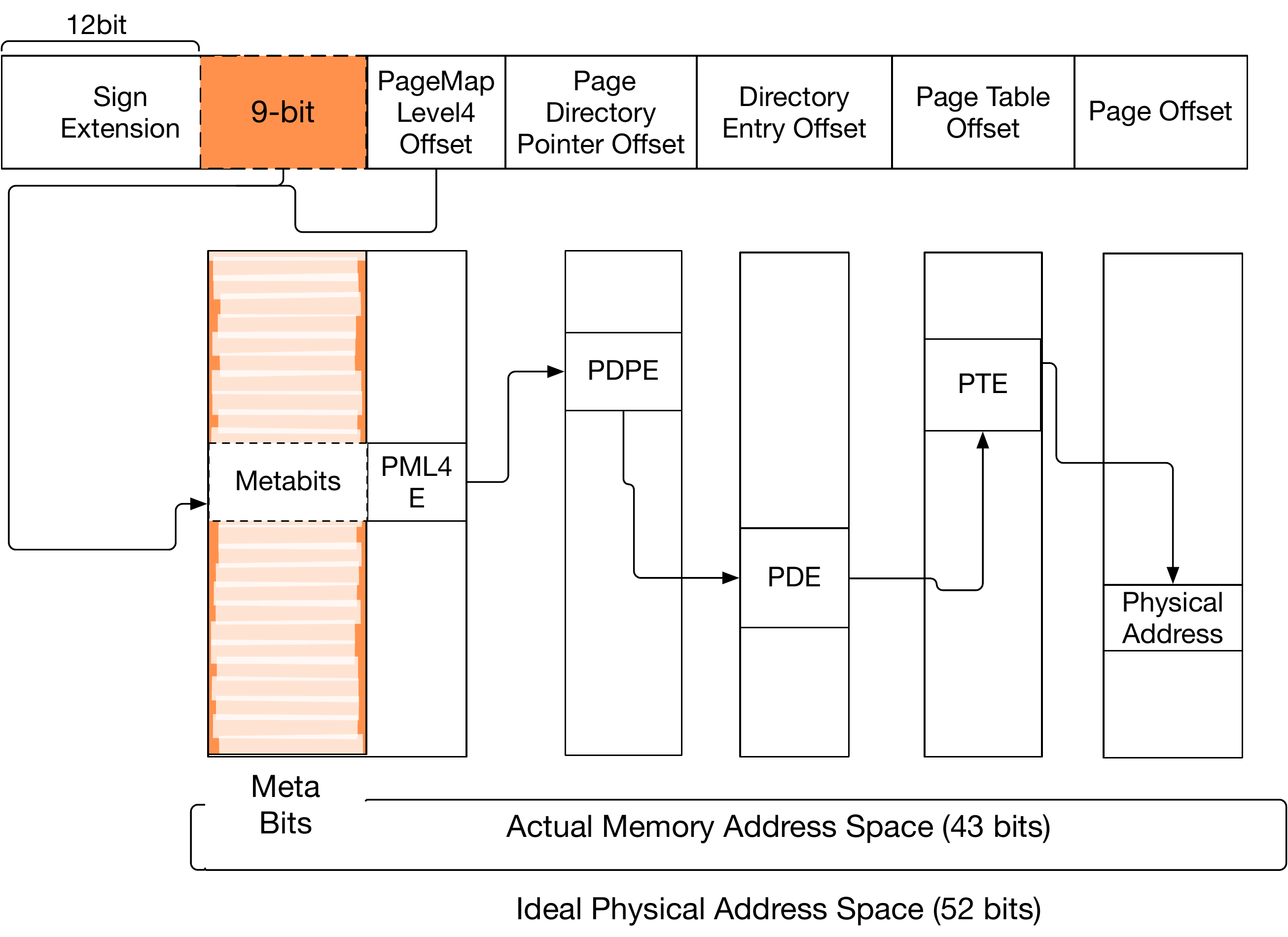}}
    \caption{Encode monitor metadata information in physical address}
    \vspace{-1em}
    \label{fig:Metabit}
\end{figure}

To answer to the first question, we encode monitoring metadata into the physical address. We use a few bits (9 to be precise) in the physical address that are only reserved by us for this design. We call them {\tt Metabits} and place them in some of the highest bits of a physical address as in Figure \ref{fig:Metabit}. We made such bold use of highest bits on a physical address based on one fact. That is, although current physical address space can be extended up to 52 bits and the real implementation from AMD supports 48 bits\cite{AMDPAE}, rare servers use memory space up to 256TB \cite{AMDPAE}. Therefore, the highest bits are free to use (The scheme to set the metabits will be discussed later), but at the cost of supporting smaller physical memory, say, only up to 8PB. Simply, our design is to modify the minor page fault handler so that if the virtual address causing a page fault comes from a DAX-mmaped file, a virtual address-specific bit segment will be written to the \textit{Metabits} of its physical address at the virtual-physical mapping stage. 
To answer to the first question, we encode monitoring metadata into the physical address. We use a few bits (9 to be precise) in the physical address that are only reserved by us for this design. We call them {\tt Metabits} and place them in some of the highest bits of a physical address as in Figure \ref{fig:Metabit}. We made such bold use of highest bits on a physical address based on one fact. That is, although current physical address space can be extended up to 52 bits and the real implementation from AMD supports 48 bits\cite{AMDPAE}, rare servers use memory space up to 256TB \cite{AMDPAE}. Therefore, the highest bits are free to use (The scheme to set the metabits will be discussed later), but at the cost of supporting smaller physical memory, say, only up to 8PB. Simply, our design is to modify the minor page fault handler so that if the virtual address causing a page fault comes from a DAX-mmapped file, a virtual address-specific bit segment will be written to the \textit{Metabits} of its physical address at the virtual-physical mapping stage.

\begin{figure}[htbp]
    \centerline{\includegraphics[scale=0.20]{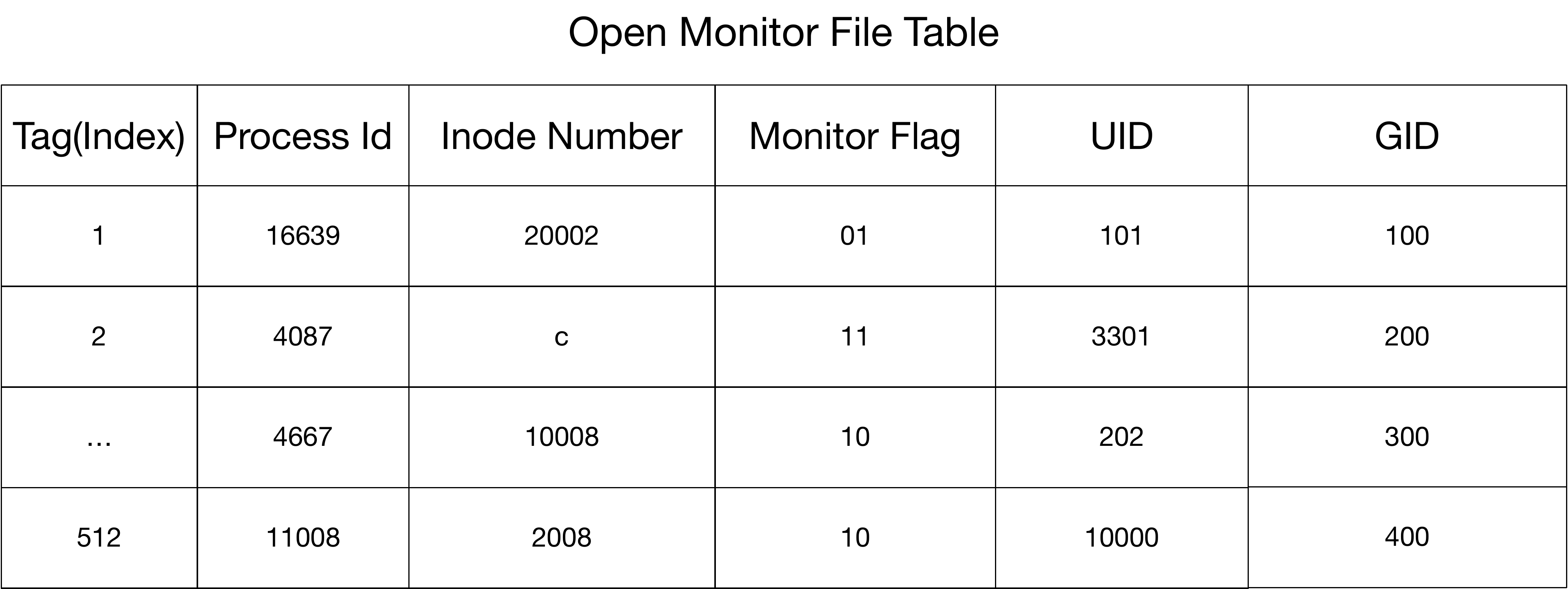}}
    \caption{Open Monitor File Table}
    \label{fig:openmonitor}
\end{figure}

\begin{figure}[htbp]
    \vspace{-0.8em}
    \centerline{\includegraphics[scale=0.22]{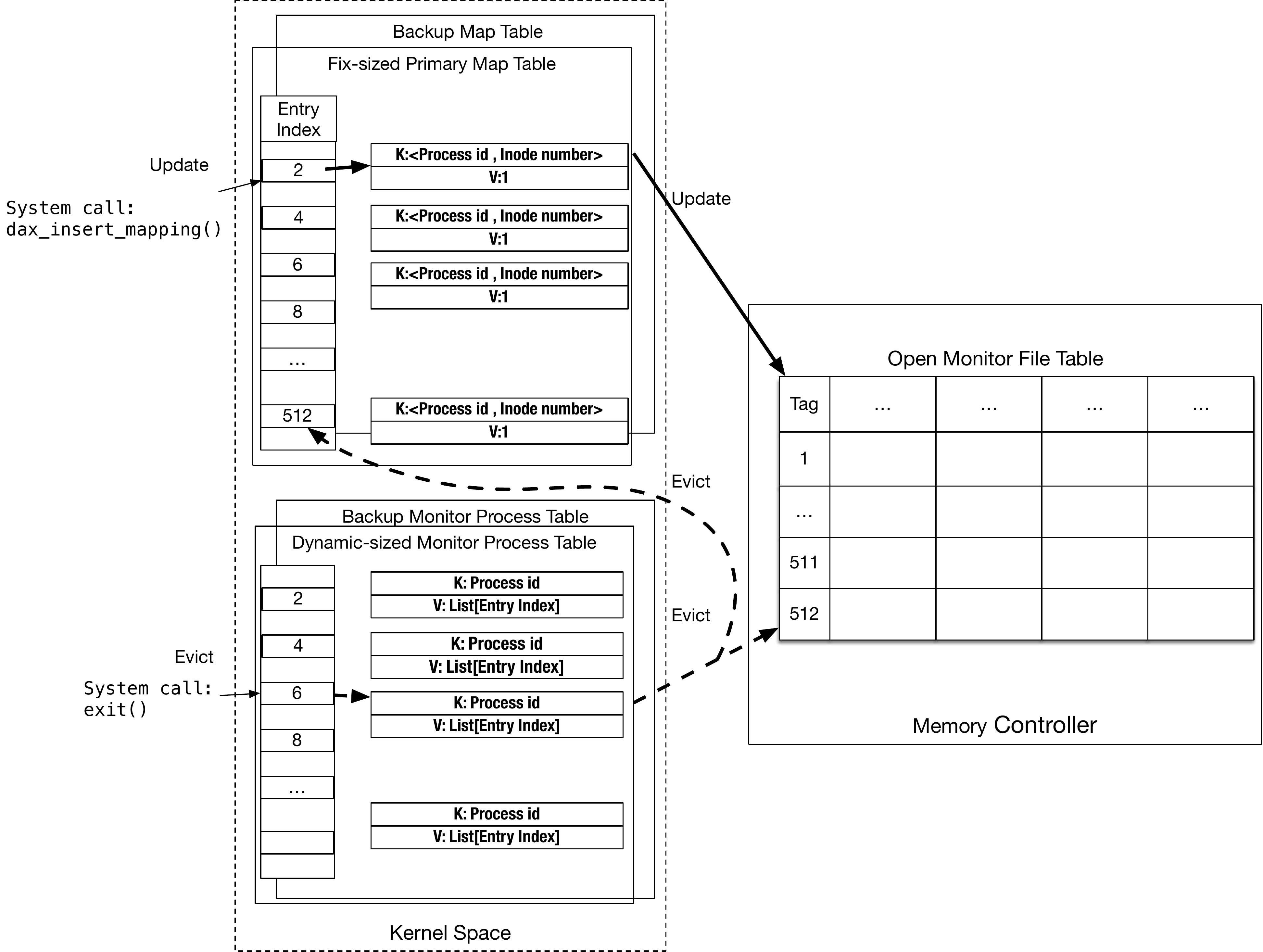}}
    \caption{How to update the Open Monitor File Table using two software tables}
    \vspace{-1em}
    \label{fig:openmonitor2}
\end{figure}

\indent{} The next question would be how and when to communicate monitoring metadata to hardware (memory controller), not only including the \textit{Metabits}, 
but user ID, group ID and more.
As discussed before, current memory controller is unable to distinguish monitor metadata from a requested physical address. Hence, we add a metadata process module to identify a monitor-required address. Besides, we need to know whether the request type is compatible with  monitor flag set by the file. The monitor flag here refers to the 4 types of operations in need of monitoring, i.g, 01 means read-only-monitor, 10 means write-only-monitor and so on. 
Moreover, we need more information such as user ID, group ID, inode number and etc, as they are required for a log record. Apparently only kernel space has all this information, and thus we introduce a hardware table to store such information from kernel. The table is called \textbf{Open Monitor File Table (OMFT)}, as shown in Figure \ref{fig:openmonitor}. This table provides information about active-monitored files and their associated process id and user information. This table serves multiple purposes, and plays a key role in our monitoring design. However, due to space limit on chip, it has only 512 entries, 17 bytes per-entry, with a total size around 9K. Apparently, this table needs to be timely and accurate. To guarantee this, we first illustrate how this table is updated (Figure \ref{fig:openmonitor2}).

To enable accuracy, the entries for \textit{OMFT} all come from a \textbf{Primary Map Table(PMT)} in the kernel space that records the active process-and-file combinations in the system. This fixed-size table is implemented using a map structure, in which the \{key, value\} pair corresponds to \{\{Process ID, Inode Number\}, 1\}, and same to the Open Monitor File Table with a total of 512 entries. Whenever a dax\_insert\_mapping() is called, its inode number, the monitor flag and the process id will be identified. If the monitor flag is not set to 00(no-monitoring), the tuple, \{Process ID, Inode Number\} will be used as a key to insert into the \textit{PMT}.
Meanwhile, another software table called \textbf{Monitor Process Table(MPT)} will also be updated with a tuple composed of \{Process ID, List of [Entry Index] \}. Next, the index of the newly inserted entry in the \textit{PMT} will be calculated and all the relevant metadata regarding its associated process will be written into a hardware \textit{OMFT} entry with the same tag value. 

While the \textit{PMT} serves as a process-file combination update source table, the \textit{MPT} is to facilitate the eviction of entries in the \textit{OMFT}. An eviction take places when 
the kernel finds out a process id involved in an exit() call is present in the \textit{MPT}. Then all the associated entries in the \textit{OMFT}, the \textit{PMT} and the \textit{MPT} are deleted consequently. The reason we employ two software tables as middlemen for metadata update is multi-folded. First, they can timely reflect the process status in the system, and the search for a certain process or combination is easy via their mapping structures. Second, an entry index from one software table just corresponds to a tag in hardware table entry as the \textit{OMFT} is implemented as a direct mapping cache. However, the \textit{PMT} is size-limited and therefore the processes beyond the table size are to be recorded in a dynamic-sized \textit{Backup Map Table} and a dynamic-sized \textit{Backup Monitor Process Table}. More discussion about the handling of these processes is in Section III.G.

Additionally we add two important components: the \textbf{Metadata Processing Module(MPM)} and the \textbf{Monitor Module(MM)}(Figure \ref{fig:Fox}). The function of the \textit{MPM} is to check the monitor-relevant information for each request, to compare the request operation to the metadata, to collect monitor content from the \textit{OMFT} and to send monitor-requested address and operation type to the \textit{MM}. The \textit{MM} reads the content from an monitor log address, identifies the buffer\_index, packages the request content, and finally issues requests.

Shown in Figure \ref{fig:Fox} is the big picture of Fox design. We provided two flavors of design for secure NVM and regular NVM. In principle, there is no much difference for both schemes, both having 3 phases. Yet in secure NVM with counter-mode encryption, there is a counter fetch step as encryption/decryption is required for passing data to/out of NVM from processors. Hence there afe counter cache and AES engines in the left part of \ref{fig:Fox}. Below, our explanation of the flow-chat is based on secure NVM.

\textbf{Initialization Phase: Encode metadata to physical address and sync to hardware table}\\
The original phase starts from the OS kernel area and only be called at page fault stage for each process. At the beginning, dax file page fault will only be handled in function dax\_insert\_mapping, where it also provides the associated information regarding the process id, inode number to the \textit{Primary Map Table}, as well the \textit{Monitor Process Table}. 
During the copy of the entry to the \textit{OMFT} on-chip 
However, since the page has never been called before, a minor page fault will be sent to the page fault handler 
where it checks the associated process id in the \textit{PMT}. If a hit is found, during the physical-to-virtual address mapping, the index of that entry is to be written to the \textit{Metabits} of the physical address to mark that the address is from an monitor-requested file(like in Figure \ref{fig:openmonitor2}). Next the physical address will be passed to the MMU to continue the request.\\

\textbf{Monitor Phase 1: Identify monitor metadata}\\
Once the application request (assume a write request) goes through the LLC, it will be handled via the regular processing path to get its data block. When it is the time to fetch the counter associated with the data block, the request address will be sent to the \textit{MPM} into a \textit{Counter Request Queue} to wait for counter address being calculated. 
When a request entry is popped from the queue, its address is saved in a register called A and its \textit{operation\_type} is saved in a bit called O, appended by another register called S. As designed, the request addresses have two types: 1.) From an address with \textit{Metabits} 2.) From an address with no such bits. While the calculated counter address is sent to the counter cache/NVM to get its value, if the address in Register A is from type 2, then the request is not monitor-requested and 0 is written to S. However, if the address in Register A is from type 1, where the \textit{Metabits} do not equal to value 0, \textit{MPM} will checks its monitor flag 
in the \textit{OMFT} based on the \textit{Metabits} and saves the whole line from \textit{OMFT} in the register S. 
If the request address contains monitor metadata, 
the \textit{operation\_type} of the request will be compared with the monitor flag previously save in register S. If they are a match, it means an monitor request is identified. 
At this moment, the request address in register A, the operation type, as well as the content in register S in combined with the current TIMESTAMP will be passed to the Monitor Module to keep a record of monitor event in the buffer log.\\ 
\textbf{Monitor Phase 2: Write to a minitor log}\\
 When the requested address and all the relevant monitored record data arrives at the \textit{Pre-monitoring Queue(PAQ)}in the Monitor Module, an monitor record is to be written to a log file. Since all the monitor items for one request is not even of a cacheline size, the \textit{MM} needs to perform a read-before-write scheme for any monitoring request. This scheme requests a read to an monitor log block and its associated counter block first before a real monitoring write. When an entry from \textit{PAQ} is popped, its address 
 will be stored in the register X, the rest content after some logic processing will be stored in the register M.  
 Meanwhile, the monitor log address is being calculated as well. Then the data and the counter for that log address are fetched from NVM or/and counter cache and together are turned to a plain-text that is to be stored in a tiny \textit{Monitor Cache (MC)} for further modification. 
 The contents from register X and register M will be combined together to be written into one cacheline in \textit{MC}  
 Finally, the real monitor write request, fully packaged, is passed and handled in memory controller. Till now, an monitor request is finished.\\
\textbf{Multiple-Tenant File Sharing}
\indent{} We further discuss here about the advantage of using Metabits in physical address other than to filtering out the non-monitor-requested addresses. Actually, when two or more different processes are mmapped to the same monitor-requested file using shared mode, if without using \textit{Metabits}, these processes will modify the same addresses in the cache and it is unlikely to know which process does the modification without any hint.
 
 As discussed before, our solution maps the \textit{Entry Index}s of process-file entries into the \textit{Metabits} to differ the processes accessing the same file, and cheats the cache that these processes intend to reach different addresses, even though the same valid addresses at memory level are obtained after trimming off the \textit{Metabits}. 
Note that this approach is necessary when MAP\_SHARE is used among different processes. The processes that share the same file using MAP\_SHARE have to msync every update to the memory, or use semaphores or other locking schemes to control the access to maintain the cache coherency. But these are known practice for file-sharing and multiple-threaded applications. Our approach does not complicate the procedures. With our approach, multi-processes/multi-threaded processes can collaboratively modify the same file without losing monitoring records. 

\subsection{Persist data-monitor ordering}
To accurately record monitor history, we need to maintain correct orders of memory requests and monitor requests. Although file operation monitor in NVM is not necessarily as strict as NVM logging, it has to follow write ordering \cite{Efficientlogging}. First, an monitoring request should be strictly after its leading data request or memory access. Second, monitoring requests themselves should follow a correct order. To this end, we locate coupled memory access and monitor request at the same memory controller. Moreover, we use \textit{MPM} to process counter request for each application request and by then to check address metadata determine monitor issue. Memory request counter is process is in parallel or ahead of an pre-monitor request which is sent to the \textit{PAQ} in the Monitor Module. The order of request and monitor hence can be guaranteed.

\subsection{Format of monitoring Logs}

To identify which information are generally expected from monitoring logs, we studied the attributes commonly used in monitoring tools and modules \cite{osquery,sysdig}. 
While our scheme can be easily extended to include more attributes, we focus on monitoring the major attributes used in state-of-art schemes and expect them to be sufficient for most use cases of monitoring trails. Finally, monitoring attributes should be unique and use notions that can be interpreted by OS and/or monitoring modules. Table \ref{tb:monitorItems} shows the main attributes on each monitor log. 

\begin{table}[h]
\center{}
\label{tb:monitorItems}
\caption{monitored Items}
\begin{tabular}{ll}
\hline
Field &
Length \\
\hline 
monitor Block Address & (max. of 8 bytes)
\\
User ID  &  (max. of 4 bytes)
\\
Group ID &  (31 bits)
\\
Action Operation &  (1 bit)
\\
Action Timestamp &  (8 byte)
\\
monitor File Inode &  (4 byte)
\\
monitor Directory* &  (4 byte)
\\
Process ID* &  (max. of 2 byte)
\\
\hline
\label{tb:monitorItems}
\end{tabular}
\end{table}



\textbf{Monitored block address} simply records the address of the transaction being monitored, i.e., the address we are reading/writing from/to. In the worst case, we have to record the full address, 64-bit address. However, if the monitoring space is pre-allocated and each monitored memory location maps to fixed monitoring region, then the full address can be inferred from the monitor location along with part of the full address. 
\textbf{monitor operation} can be as simple as read or write. 
\textbf{User ID} and \textbf{Group ID} provide information for access authority record. \textbf{Timestamp} gives accurate time access for history trace. \textbf{Inode ID} can be used to infer the file name. 
If security level is high, more information can be recorded. Then the monitor entry could be switched to a rich-content mode, with the optional items. However, we leave to the users to make the decision.

\subsection{Monitoring storage schemes and overhead}

Since memory controller will be responsible of writing monitor logs, it should also know the address of where to write such logs to. Thus, a trade-off between design complexity and storage efficiency is expected. For instance, a scheme that prefers design simplicity over storage overhead would pre-allocate part of the memory for monitoring trails, we call it \textbf{fixed-location scheme} or storage-oriented scheme. For instance, for each 64B memory block there is a fixed space, e.g., 32B, allocated for monitoring logs for accesses targeting that memory location. All the monitoring buffers are located after usable memory range, e.g., the last 32GB of a 96GB memory are used for monitoring. However, in such a scheme, the memory controller can quickly locate the monitor buffer address from the memory access address. Clearly, such a scheme has significant storage overheads when used with fine granularity, e.g., 64B blocks, and also have limited history; only tracks the last memory access to a specific location, which does not meet the standard monitoring requirement - temporal history is mandatory for monitoring \cite{CC}.

An alternative scheme would be pre-allocate a large circular buffer that has its index tracked in the memory controller. We refer to such a scheme as \textbf{global circular buffer} orhttps://www.overleaf.com/project/5de4825c35a49a0001a3305b history-oriented scheme.  In global circular buffer, after each memory access, the index (pointer) of the buffer is incremented (mod-ed by buffer size) to point to the next location to write the next monitor log to. However, such a global buffer scheme would lose the high resolution provided by spatially-focused buffers as in fixed-location scheme. For instance, if file A is being accessed heavily while another one, file B, is also accessed but less often, then the global buffer would contain little and possibly no information about accesses to file B. In other words, the history of frequently accessed files would dominate the monitor logs and history of less frequently-accessed files will get erased.


In that case, we can reserve a global monitor circular buffer that may use roughly 1/20 of the whole memory space for practical concern. In case a history of up to three-month to half-year of monitor records is requested, the global buffer can provide records for all monitored files. A counter could be set for this buffer to count the write number to the buffer. When a certain percentage, i.e 50\% of the global buffer is reached these records could be backed-up to a secondary storage, or be backed-up daily to avoid record being overwritten via a small application.

One important aspect that we have to carefully consider is the granularity of monitor writes. Memory systems read and write data with 64B granularity. However, as our monitor entries are smaller than 64B, we need to either expand entry size to 64B or simply read the whole 64B around the monitor log, update the part of the monitor entry, and then write the whole 64B back to memory. The former scheme is impractical as it can easily double or triple the storage overhead of monitor logs, while the later scheme would incur a memory read before writing each monitor log. Fortunately, we observe a high temporal and spatial locality of monitor blocks, and thus by caching them we can skip the memory read part. 

\subsection{Monitoring schemes}
It this part, we will discuss three schemes with their trade-off.  
\textbf{Full monitoring scheme}
The most straightforward and protective scheme is the full monitoring, where each read/write memory access will be monitored at the memory controller.  This is the control scheme in our paper due to huge overhead. 
\textbf{Selective monitoring scheme}
This is the major monitoring scheme we devise for this work, where only dax-mmapped region, or certain access type, or a combination of both to be monitored. The monitoring writes will be reduced proportionally according to the ratio of selective address range to the whole memory size and to application-specific behavior. \textbf{Directory-based monitoring scheme}
Further we provide an monitoring scheme that is similar to software or system level monitoring option. Widely seen in market, one or multiple folders are assigned by users for specific monitoring/encryption purpose\cite{DirectoryAudit}. In this scheme, we pass the user-defined directory name as a parameter to kernel to enable directory-based monitoring. With this scheme, any file that is created and only under this directory is given monitoring permissions by Fox. 

\vspace{-1.5em}
\subsection{Discussion}
First, we should address our current schemes has its limitation.
If process-file combinations are more than 512, we could not assign 
an unique index to distinguish them, because we can only re-purpose up to 9 bits in the physical address to indicate monitor-able addresses. The rest combinations therefore will not have any marked addresses during the page fault stage, and therefore will only be recorded in the backup table and the backup process table. To save a record, we employ the global circular buffer to record the mmap() and exit() information for these combinations. 

However, we argue that actually the number of file on a single system that requires monitoring is usually below 100 in business setting, and processes usually rapidly access them and exit. Furthermore, classified files usually have strict access control. If read only, one cannot mmap (PROT\_WRITE) them at all. Therefore, our design still is practical at present and in near future, which could be also employed on distributed systems. But, for high throughput system, such as OLTP database, optimization is required for our scheme. Our design fits very well for OLAP database, managerial systems, personal computer, or those system that have highest security requirement.

\section{Methodology}
\label{sec:method}
In this section, we provide the details about the simulation and evaluation methods used in this study.\\
\noindent \textbf{Architecture Simulation:}
We use Gem5 for modelling and simulation \cite{GEM5}. The simulation is performed using full system mode, with a ubuntu kernel of version 4.40 and a 16.04 ubuntu image. The architecture configuration parameters are shown in table \ref{tb:config}.We implement a 256KB, 16-way set associative counter cache, each with a total number of 4K counters for file encryption and memory encryption respectively. To stress-test our design, we select memory-intensive applications. Specifically, we select 3 memory-intensive representative benchmarks from the SPEC2006 suite \cite{SPEC}, 3 persistent micro-benchmarks implemented with PMEM library\cite{PMP} or Intel persistent instructions and 2 Whisper persistent benchmarks \cite{WHISPER} for different experiments in this work. Each benchmark accesses a single file, which is either allocated on NVM or on a regular block device, and all the intermediate results will be written to or read from that specific file. The SPEC2006 applications serve as control groups for NVM-based and directory-based monitoring.

Additionally, we generate 8 mixed benchmarks consisting of both persistent and non-persistent applications.  The goal is to evaluate the performance and the overhead of persistent and non-persistent benchmarks on our current design and the impact of their interplay on performance. To accurately record page fault number from persistent benchmarks, we store fault addresses in the pre-allocation stage before checkpoint and restore them before continuing with 200 million instructions.  Non-persistent benchmarks are fast-forwarded, followed then by the checkpoint after 2 billion instructions(4 billion for mixed benchmarks) at the full system mode. 
Similar to prior work \cite{OBFUSMEM}, we assume the overall AES encryption latency to be 24 cycles, and we overlap fetching data with encryption pad generation. For all the experiments, we use balanced scheme as our storage plan. However, we think the performance overhead of our results should be similar to the global circular buffer scheme.  
\vspace{-1em}
\begin{table}[htp!] 
\centering
\caption{Configuration of the simulated system.}
\label{tb:config}
\scriptsize
\begin{tabular}{|l|p{5cm}|} \hline
\multicolumn{2}{|c|} {\bf Processor} \\ \hline
CPU & 8-core, 1GHz, out-of-order x86-64 \\
\hline 
L1 Cache &  private, 2 cycles, 32KB, 2-way, 64B block\\
\hline
L2 Cache &  private, 20 cycles, 256KB, 8-way, 64B block\\
\hline
L3 Cache & shared, 32 cycles, 8MB, 64-way, 64B block\\
\hline
\multicolumn{2}{|c|} {\bf DDR-based PCM Main Memory} \\ \hline
Capacity & 16 GB\\ 
\hline
PCM Latencies & 60ns read, 150ns write \cite{PCM2} \\
\hline
Organization & 2 ranks/channel, 8 banks/rank, 1KB row
\newline
buffer, Open Adaptive page policy, RoRaBaChCo 
address mapping\\
\hline
DDR Timing & tRCD 55ns, tXAW 50ns, tBURST 5ns, tWR 150ns, tRFC 5ns \cite{PCM2,CrashC}\\ 
& tCL 12.5ns, 64-bit bus width, 1200 MHz Clock\\
\hline
Persistent Memory & 4G, from 12GB to 16GB \cite{PMP}\\
\hline
\multicolumn{2}{|c|} {\bf Encryption Parameters} \\
\hline
Memory Counter Cache & 128KB, 8-way, 64B block\\
\hline
\multicolumn{2}{|c|} {\bf Monitor Parameters} \\
\hline
Monitor Cache & 64KB, 8-way, 64B block\\
\hline
\end{tabular}
\end{table}

\noindent \textbf{Kernel Modification:}
We made the following major modifications to Linux kernel 4.14. 
In dax.c we read and write mapped physical addresses caused by minor page faults in \textbf{\textit{dax\_insert\_mapping}} function. To obtain the requested file's Inode Number as well as the requested Process ID , we output them from \textbf{\textit{dax\_insert\_mapping}} function into different ide registers in Gem5. Moreover, in dax.c, we implement a function called \textit{Getfullpath} to identify whether a request is from a file under monitor-requested directory, and output the result to a register in Gem5. Both software tables aforementioned are implemented in dax.c file and be shared by exit() system call.\\ 
\noindent \textbf{Schemes:}
In this paper, we compare following schemes:
\begin{enumerate}
    \item \textbf{Full Non-volatile Memory Encryption:}
    This is one of the baselines for our study. With this scheme, no monitoring write is issued. Each memory access incurs decryption/encryption overhead. \item \textbf{Full Non-volatile Memory Monitoring with Encryption:}
    This is another baseline in our study. In this scheme, each memory access incurs a memory write.
    \item \textbf{Address-selective Non-volatile Memory Monitoring with Encryption:} With this scheme, only files within certain addresses will be monitored.
    \item \textbf{Operation-selective Non-volatile Memory Monitoring with Encryption:} Under this scheme, monitoring operation is restricted to certain memory access type.
    \item \textbf{Directory Monitoring with Encryption:} Under this scheme, only files under monitor-requested directory will be Monitored. 
\end{enumerate}
\subsection{Benchmarks}
We used 2 Whisper applications, 3 PMDK micro-benchmarks, and 3 memory-intensive SPEC2006 applications to study file monitoring in this paper. The combinations of all the benchmarks are listed in the table \ref{tb:mix} below as well. For any combined benchmark we pin each application to a separate core during experiments to avoid CPU resource conflicts.
\begin{itemize}
\item{\textbf{Swaparray}}: Insert and swap random items in a persistent array. Transaction persistency is guaranteed via PMDK.\\
 \vspace{-1em}
\item{\textbf{Hashtable}}: Insert and remove random items in a persistent hashtable. Transaction persistency is guaranteed via PMDK.\\
 \vspace{-1em}
\item {\textbf{Daxbench}}: Mmap a persist file of a size taken from a parameter and read/write it with a ratio taken from a parameter. Persistency is guaranteed via clflush and sfence.\\
  \vspace{-1em}
\item{\textbf{Btree}}: Insert random items and search random items in a persistent balance tree. It's a demo program from PMDK.\\
  \vspace{-1em}
\item{\textbf{Hashmaptx}}: Insert random items and search random items in a persistent hashmap. It's a demo program from PMDK.\\
\end{itemize}
 \vspace{-2em}
\begin{table}[htp!] 
\centering
\caption{Mixed Benchmarks}
\label{tb:mix}
\scriptsize
\begin{tabular}{|l|p{7cm}|} \hline
\multicolumn{2}{|c|} {\bf Benchmark Combination} \\ 
\hline 
1 &   MCF, Swaparray\\
\hline 
2 &   Btree, Hashtable \\
\hline 
3 &   LBM, LIBQUANTUM \\
\hline 
4 &   DAXBench1282 Swaparray\\
\hline
5 &   Hashtable, Hashtable (Test: One from Monitor-requested directory)
\\
\hline 
6 &   Swaparray, Hashtable (Test: Hashtable from Monitor-requested  directory)\\
\hline 
7 &   Swaparray, Hashtable (Test: Swaparray from Monitor-requested directory) \\
\hline
8 &   MCF, DAXBENCH1282 (Test: none from Monitor-requested directory)\\
\hline
\end{tabular}
\end{table}
\section{Evaluation}

\label{sec:eval}
\begin{table}[]
\centering
\caption{MPKI and IPC}
\label{tb:mix}
\begin{tabular}{lll}
\hline
Application & MPKI & IPC  \\
\hline
Btree       &0.84 & 0.1474 \\
\hline
DAXBENCH128   &50.56 & 0.154\\
\hline    
Hashmaptx   &  0.81 & 0.2428 \\
\hline
LBM         &  28.49& 0.295 \\
\hline    
Swaparray   &  2.99 & 0.880 \\
\hline    
LIBQUANTUM   & 13.24 & 0.769 \\
\hline    
Hashtable   &  3.03  & 0.658 \\
\hline
MCF        & 55.31 & 0.260 \\
\hline    
\end{tabular}
\end{table}
In this section, we describe the evaluation results we obtained from our design as well as the relevant analysis based on these results. Our results validate the design we implemented. 

\subsection{Results and Analysis}
First of all, we list the MPKI and IPC statistics for all the applications we used in our evaluations in Table IV.

\begin{figure}[htbp]
    \vspace{-0.8em}
    \centerline{\includegraphics[angle=270,origin=c,scale=0.38]{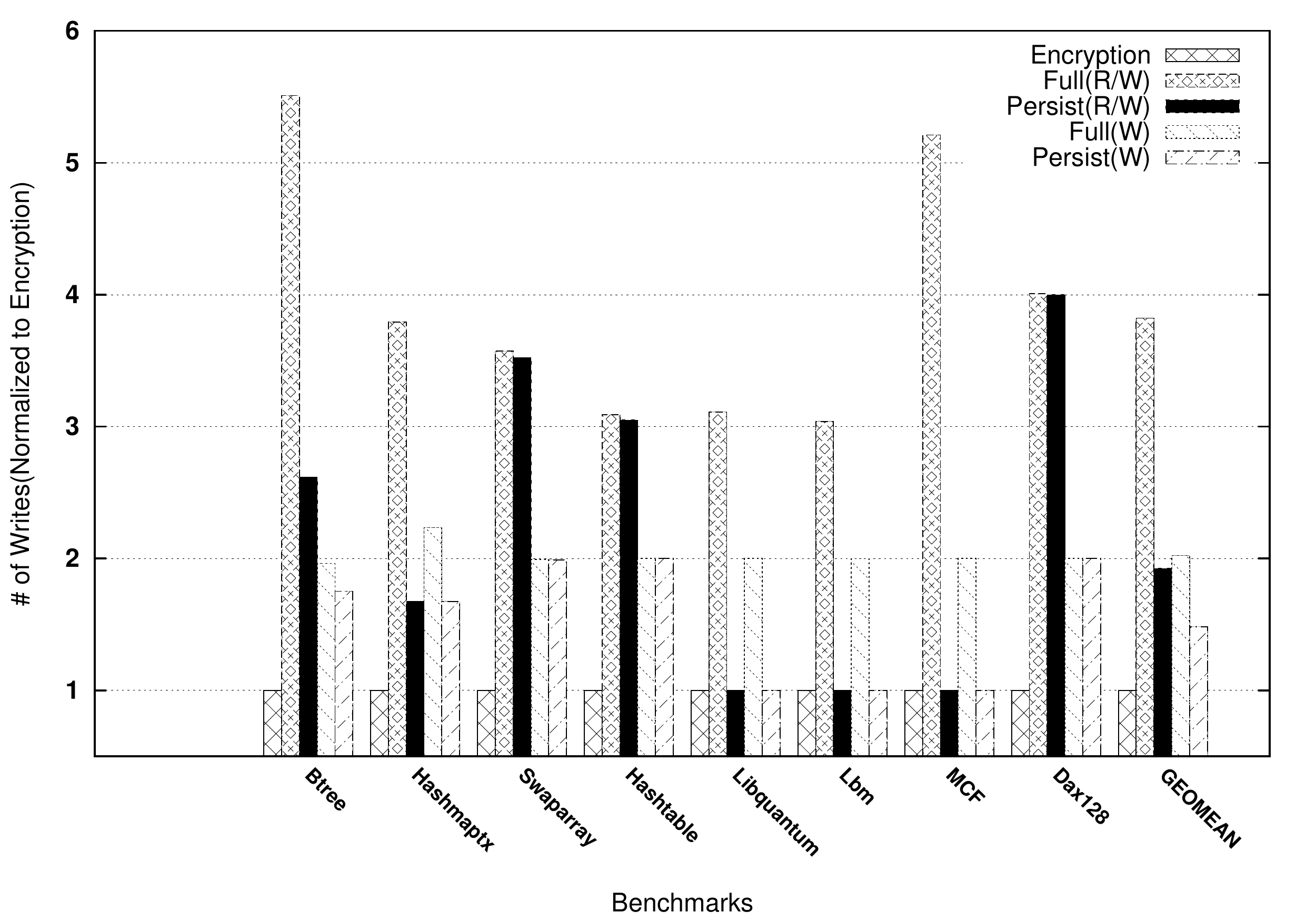}}
    \vspace{-4em}
    \caption{The impact on number of writes for different schemes}
    \label{fig:write}
    \vspace{-1.5em}
\end{figure}
In Figure \ref{fig:write}, we sketched the number of writes normalized to the no-monitoring encryption baseline. Thus bars labeled with Encryption are of value 1. For the Full(R/W) scheme, as aforementioned in motivation part, this scheme is the most detailed, but the write overhead is too large to bear, ranging from 3 to 5.5 times of the baseline for all benchmarks. Since we use three different groups of benchmarks, their behaviors are quite distinguishable and have their own signatures in these schemes. For whisper applications(Btree and Hashmaptx), they both show a 50\% reduction of writes when only requests that hit their pre-allocated NVM pools are monitored in the Persist(R/W). This is in consistent with how these applications are implemented as noticed by  \cite{DudeTM}, that quite a number of requests are to maintain metadata of transactions in random memory areas.  For microbenchmarks Swaparray, Hashtable, and DAXBenchs, on the opposite, nearly all their requests hit pre-allocated memory pools. Therefore, only slight reduction of writes are observed in the Persist(R/W) scheme. As expected, all applications from SPEC2006 do not hit the emulated NVM region because they are not persistent applications. Thus using selective monitoring schemes targeting persistent memory does not incur any extra writes. In write-only monitoring schemes, the number of writes is really application-specific for each application.
Overall, the write-only monitoring schemes significantly reduce the write overhead, by 46.4\% (full memory) and 61.3\%(selective area) compared to Full(R/W) scheme, and incur only 48.2\% extra writes compared to no-monitoring scheme.

\begin{figure}[htbp]
    \vspace{-0.8em}
    \centerline{\includegraphics[angle=270,origin=c,scale=0.38]{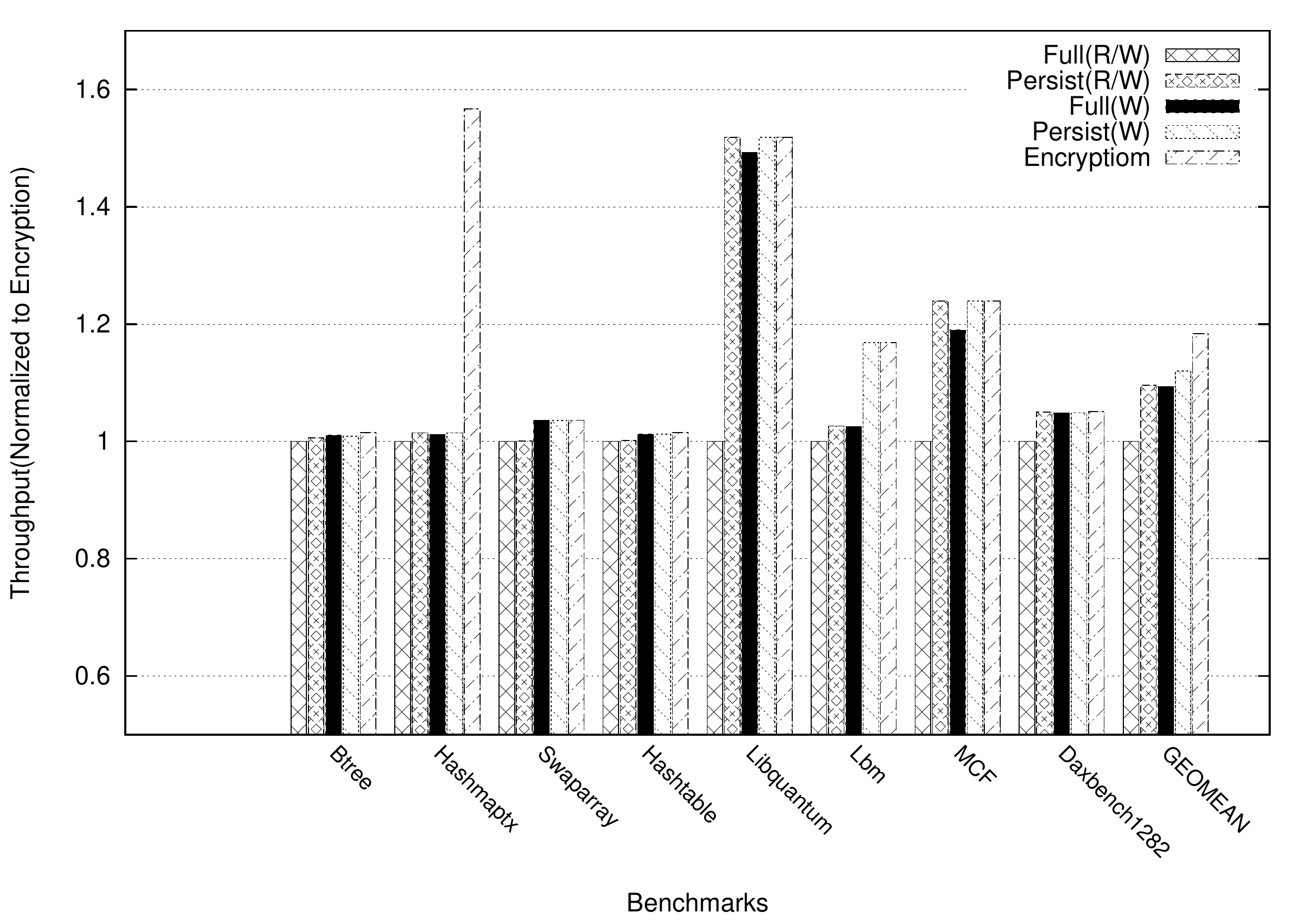}}
    \vspace{-4em}
    \caption{The impact on ipc for different schemes}
    \label{fig:throughput}
    \vspace{-1em}
\end{figure}

Next, we study the impact of different monitoring schemes on the throughput of the system. The lower boundary in this study is the Full(R/W) scheme, which theoretically consumes the most throughput among all the schemes and the upper boundary of throughput for each application is its no-monitoring encryption model as shown in Figure \ref{fig:throughput}. As expected, on average, 19\% throughput was reduced when Full(R/W) is employed compared to the no-monitoring scheme. Persist(R/W) and Full(W) schemes act similarly, both negatively affecting about 9\% throughput of the upper boundary. Lastly Persist(W) scheme only hurts 6.5\% of the throughput and is the most optimized scheme as designed. Therefore, throughput is slightly impacted by selective monitoring scheme. 

\begin{figure}[htbp]
    \vspace{-0.8em}
    \centerline{\includegraphics[scale=0.45]{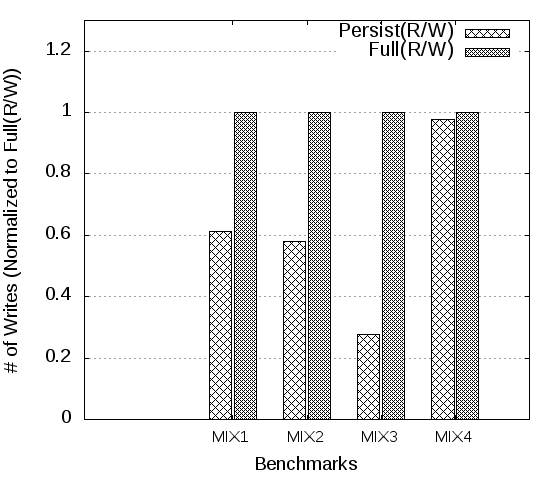}}
    \vspace{-1em}
    \caption{The impact on number of writes for Persistent/Full mixed scheme }
    \label{fig:fullper}
    \vspace{-1em}
\end{figure}

To further evaluate the interaction between applications under different monitoring schemes, we performed the experiments on mixed benchmarks. As shown in Figure \ref{fig:fullper}, we examined mixed benchmarks either using baseline scheme(Full(W/R)) or the selective version(Persist(W/R)). Theoretically, if two applications equally divide the whole simulated instructions in the both Full and Persist schemes, we can use an equation to get an estimated reduced percentage for the Persist scheme. However, in reality, applications compete for resources and non-determinism plays a role in performance. MIX1 consists of one persistent and one non-persistent application, and when persist scheme is used, the number of writes is reduced by almost 40\%.  MIX2, on the other hand has two persistent applications. But these two applications show quite different memory access pattern, an intensive and a relatively non-intensive one. Consequently, their persistent performance acts 57\% of their Full scheme.
MIX3 is made of two non-persistent applications. Therefore, when we restrict monitoring to areas that only belong to the emulated persistent memory, no monitoring id supposed to occur. As a result, the number of writes drop significantly, about 73\%, compared to the Full(R/W) scheme. A big contrast in shown in MIX4, where in both schemes they show a similar read/write pattern. The reason behind is that both persistent applications in this benchmark show a close performance in the Full and Persist scheme individually. 

\begin{figure}[htbp]
    \vspace{-0.8em}
    \centerline{\includegraphics[scale=0.45]{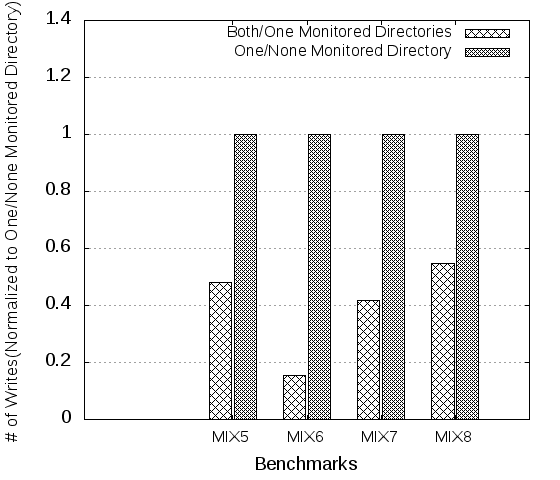}}
    \vspace{-1em}
    \caption{The impact on number of writes by Directory-restricted/Directory-free mixed scheme }
    \label{fig:Directory}
    \vspace{-1em}
\end{figure}

Our most fine-grained monitoring scheme is directory-based scheme. Only the files created under an selected directory are monitored. Again, we tested mixed benchmarks to evaluate this scheme. As shown in Table III, MIX5 is made of two same persistent applications but one is with an monitor-requested file and the other is not. MIX6 and MIX7 use two different persistent applications, both showing memory-intensive behaviors. MIX 8 contains one persistent and one non-persistent applications. Consistent with our expectation, in Figure \ref{fig:Directory}, for MIX5, when only one application accesses the file under monitor-requested directory, the number of writes is reduced by almost half compared to the case that both applications can access monitor-requested files. Similarly, MIX6 uses two micro-benchmarks(Hashtable and Swaparray) However, it reveals that the total writes for only one process (Hashtable) handling monitor-requested file in persistent memory leads to a significant write reduction to only 15.7\% of that from the two monitor-requested file case. We further did one experiment for the same combination from MIX6, but exchange the file directory for each benchmark, shown as MIX7 in this figure. This, however, makes a write number reduction to 41.7\% of the dual monitor-requested-file case. From these results, we can make the conclusion that Swaparray has many more read access during evaluation period than Hashtable, although both are memory-intensive applications. The reason behind is that read instructions have to be processed quickly but writes are reordered and delayed if no-dependence between them. We can see dramatic write reduction caused by fast processing of non-monitor-requested Swaparray, which infers Swaparray has a lot of read meanwhile. On the opposite, when Swaparray access monitor-requested file, the writes maintains a high percentage, indicating it is not impacted by Hashtable that much. 

Interestingly, for MIX8, of which one application from SPEC2006 has no access to monitor-requested files and the other a persistent benchmark but does not deal with monitor-requested file either, shows only 57\% of the writes of that when its persistent application handles an monitor-requested file. So compared to MIX6 and MIX7, although all the writes come from regular requests, the non-monitor-requested experiment of MIX8 still makes a high percentage of writes. 
In all, directory scheme works for all our mixed benchmarks.

Additionally, we did the sensitivity studies on monitor Cache size, and we found that when varying cache size from 8K, 32K, 64K, up to 256K, starting from 64K the execution time does not decrease any more. So we choose 64K as a final parameter in our evaluations.


\section{Related Work}
\label{sec:rela}
In this section, we review and discuss the relevant work in NVM-supported filesystesm, and hardware-supported logging. \\
\noindent \textbf{NVM-supported filesystem} \\
Currently there are two types of filesystems which support the integration of NVM. One type is DAX-supported legacy filesystem \cite{XuThesis}. The other is the NVM-specific filesystem \cite{NOVA,xu2017nova,pmfs,HiNFS,condit2009,NVMSEFS}. The DAX-support filesystem include Ext4-DAX\cite{DAX, DAXforLinux}, Microsoft-DAX\cite{WindowsDAX,WindowDAXproblem},which are the mainstream filesystem in the market. While the NVM-specific filesystem or NVM in hybrid are still at research phase\cite{xu2017nova,pmfs,HiNFS,condit2009,NVMSEFS}. Nonetheless, NVM is still considered a promising candidate for secondary storage\cite{HiNFS}. PMFS is a lightweight file system that support DAX and use journaling to store metadata\cite{pmfs}. HiNFS uses direct access to support NVM storage but improves the performance of direct access write by providing a NVMM-aware write policy to hide long latency\cite{HiNFS}. BPFS does not support mmap\cite{condit2009}, SCMFS and Aerie focus on reducing the software overhead on NVM-based system\cite{SCMFS,Aerie}.
Nova is a recent log-structure file system that is purely designed for NVMM characteristics, completely avoids block layers, and works directly with storage mapped into the kernel's address space\cite{xu2017nova}. Its function-enhanced version Nova-fortis \cite{NOVA} added more functions to make it more practical as currently mainstream block-device supported filesystem. 
In \cite{kateja2019lazy}, the dirty bit in page table is repurposed to mark data redundacy and it is possible to use it to record read and write for DAX-mmaped page, but we do not take this approach in concern of overhead cost. Our works targets at dax-supported, DRAM-NVM hybrid filesystem.


\noindent \textbf{Hardware-supported logging for persistent memory}\\
Hardware-supported auditing for NVM is not an area that has been studied much. However, hardware-supported logging for NVM has been intensively studied recently\cite{atom, hardwarelog, Steal, Picl, ShadowPaging, Efficientlogging}. Atom\cite{atom} is the first piece of work that provides a complete design of hardware logging at cache and memory controllers for persistent memory. 
Proteus\cite{hardwarelog}, further proposes two new instructions to facilitate hardware logging and achieves the flexibility of software and the low overhead of hardware. More recent work \cite{Efficientlogging} compares the aforementioned papers as well as other peer work \cite{WRAP, Steal} in hardware logging and provides a undo and redo approach which efficiently reduces the log writes to NVM by asynchronous and direct data updates with the help of small reserved region in DRAM cache.
Although these papers propose solutions for a different problem, many aspects they addressed share the similar sprit in our work.\\ 
\noindent \textbf{Auditing and casualty analysis}\\
System logs and auditing logs has been long served for casualty analysis\cite{ma2015accurate,lee2013loggc,xu2009detecting,gao2018aiql,ma2017mpi,gao2018aiql,liu2018towards,goel2008reconstructing,hossain2017sleuth,kwon2018mci,king2003backtracking,sitaraman2005forensic,goel2005forensix,king2005enriching,renroot,unearthing}. Many of these works  develop forensic tools to capture and analyze logs\cite{sitaraman2005forensic,goel2008reconstructing,hossain2017sleuth}. Since network attacks, advanced persistent threats are widely presented in current cyber-and-physical environment, intrusion detection is always in demand. Current focus in this research area is the fast,accurate and automatic attacks reconstruction and alert. Our work makes contribution to provide accurate and fine-grained sequences of file events for NVM-residing files.
\section{Conclusion}
We propose a novel hardware-assisted design called Fox that monitors file operations in NVM-supported filesystem. We first notice that in DAX-supported filesystems, dax-mmap is problematic for auditing and may results in failure in attacks detection. Accordingly, we provide a hardware/software co-design to provide a fine-grained file event monitoring framework to patch this security blind spot. We provide different monitoring schemes from full monitoring to selective monitoring to reduce write overhead. Moreover, we provide a directory-based file monitoring scheme that selectively monitors files under certain directories according to the secrecy level of files. In our gem5 full-system evaluation, the throughput, on average, is reduced by 9\% to 6.4\% according to different selective schemes, but it induce 48\% extra writes to the memory. Compared to the instrumentation-based software approaches, our solution is secure, complete and low-cost.
\label{sec:con}


\bibliographystyle{IEEEtranS}
\bibliography{ref.bib}
\vspace{12pt}

\end{document}